# Dissipative instability of converging cylindrical shock wave

Sergey G. Chefranov

Physics Department, Technion-Israel Institute of Technology, Haifa 3200003, Israel

csergei@technion.ac.il

## Abstract

The condition of linear instability for a converging cylindrical strong shock wave (SW) in an arbitrary viscous medium is obtained in the limit of a large stationary SW radius, when it is possible to consider the same Rankine-Hugoniot jump relations as for the plane SW. This condition of instability is substantially different from the condition of instability for the plane SW because a cylindrical SW does not have chiral symmetry in the direction of the SW velocity (from left to right or vice versa) as in the case of a plane SW. The exponential growth rate of perturbations for the converging cylindrical SW is positive only for nonzero viscosity in the limit of high, but finite, Reynolds numbers as well as for the instability of a plane SW.

## Introduction

Shock waves ("shocks" or "SW") arise in hydrodynamics, aerodynamics and many other fundamental and applied physical problems, such as the problems of inertial confinement fusion [1], supernova explosion [2] and underwater electrical explosion of wires and wire arrays [3], [4]. In cylindrical and spherical geometries, a converging shock strengthens as it propagates towards the center; it is of high importance to examine its stability [5]-[9]. For example, in [7]-[9] it is stated that a convergent spherical shock is unstable in linear approximation, when the growth rate of the disturbances, which is obtained numerically, is only algebraic and slower than exponential. The shock is stable in the case of symmetrical perturbations [5], [6], [8] for converging spherical and cylindrical shocks, but a finite region of stability has been found numerically for the case of asymmetrical perturbations. It is also known that a plane shock is always stable in one-dimensional (1D) perturbations in an ideal medium [10]-[12]. As is shown in [10], this is the result of the absence of a parameter which has the dimension of inverse time; this is necessary, but not sufficient, to realize plane shock instability. Indeed, in [13] for the case of shocks in a viscous medium, this necessary parameter with dimension of inverse time arises, but stability of 1D perturbations is stated for the weak plane shocks considered in [13] on the basis of Burgers' equation. On the other hand, for strong plane shocks the instability of 1D perturbations is stated in [14] on the basis of a generalization of the D'yakov theory [10]-[12] when viscosity is taken into account.

For cylindrical shocks, the parameter with dimension of inverse time always exists due to the finite curvature of the shock front and the compressibility of any medium with finite speed of sound. However, instability, in the limit of large stationary radius, of a converging cylindrical SW velocity may arise only if viscosity is taken into account for strong shocks, as in the case of a plane shock front, but under substantially different conditions. Here the instability of converging cylindrical shocks in a viscous medium for the case of 1D and 2D perturbations is stated.

### 1. Dispersion equations

Let us consider a converging cylindrical SW of arbitrary intensity, propagating in the directions perpendicular to the $z$ axis in the cylindrical variables $(z, r, \varphi)$.

The radial velocity of the converging shock front is $D < 0$, and $U < 0$ is the radial velocity of the medium behind the shock wave front. For simplicity, let us consider the case when it is assumed that the SW front is uniform in the direction of the $z$ axis and there are no perturbations of the velocity field component along this axis. This case allows us to simulate a converging cylindrical shock wave arising from an explosion of a system consisting of a finite number of long wires bounding a cylindrical region with an axis coinciding with the z-axis.

In this case, perturbations in the azimuthal direction can be caused by the finite distance between the wires, which determines the wavelength of the corresponding perturbation. Since the wires are assumed to be uniform along the length, this corresponds to the assumption made above that there are no disturbances along the z-axis.

Also let us consider the quasi-stationary limit when $D \approx const; U \approx const$ and it is possible to neglect the piston influence on the shock front propagation and its stability.

In this limit the equation for the perturbed cylindrical surface of the shock $R_s = R_{s0}(\tau = \varepsilon t) + r_{sa}(\varphi, t); \varepsilon \ll 1, \tau \approx const$ depends only on time and the azimuthal coordinate $\varphi$ in the form (see Fig.1):

$$r_{sa} = g(\varphi, t) \qquad (1)$$

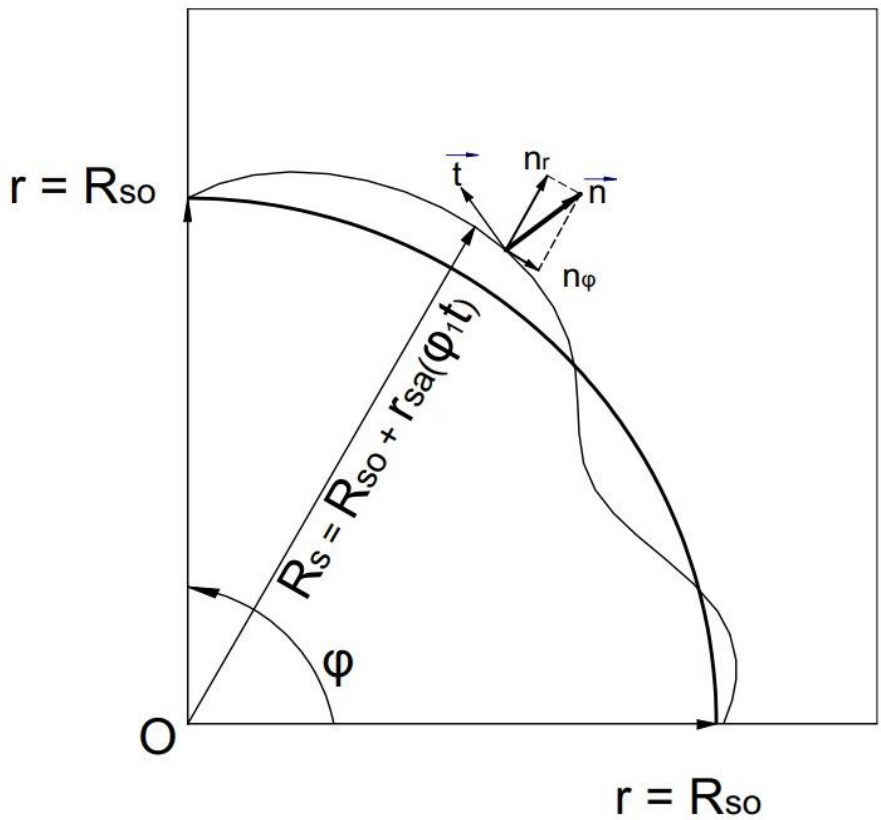

Fig.1

Caption to Fig.1

An image of a fragment of the front of a converging cylindrical shock wave is given. The unperturbed front is characterized by the magnitude of the shock wave with radius $R_{s0}$. The resulting shape of the front after the perturbation $r_{sa}(\varphi,t)$ is applied, as well as the normal vector $\vec{n} = (n_r; n_\varphi)$ to the perturbed front surface and the tangent vector $\vec{t} = (t_r; t_\varphi)$, are given. In the drawing, the point $O$ indicates the place where the shock wave is focused.

In (1), the value $r_{sa}$ characterizes a small additional radial deviation of the surface of the shock wave front from the unperturbed radius of the shock wave $R_{s0}$ when $r_{sa} \ll R_{s0} \approx const$.

Equations for small perturbations of the velocity and pressure fields behind the front of the shock wave in the linear approximation have the form in the frame of reference where the shock is at rest ($w = U - D \approx const$), when they are represented in cylindrical coordinates:

$$\frac{\partial V_{1r}}{\partial t} + w\frac{\partial V_{1r}}{\partial r} = -\frac{1}{\rho}\frac{\partial p_1}{\partial r} + \nu_1\left[\frac{1}{r}\frac{\partial}{\partial r}r\frac{\partial V_{1r}}{\partial r} - \frac{V_{1r}}{r^2} - \frac{2}{r^2}\frac{\partial V_{1\varphi}}{\partial \varphi}\right] + (\frac{\nu_1}{3}+\nu_2)\frac{\partial div\vec{V_1}}{\partial r};$$

$$\frac{\partial V_{1\varphi}}{\partial t} + w(\frac{\partial V_{1\varphi}}{\partial r} + \frac{V_{1\varphi}}{r}) = -\frac{1}{r\rho}\frac{\partial p_1}{\partial \varphi} + \nu_1\left[\frac{1}{r}\frac{\partial}{\partial r}r\frac{\partial V_{1\varphi}}{\partial r} - \frac{V_{1\varphi}}{r^2} + \frac{2}{r^2}\frac{\partial V_{1r}}{\partial \varphi}\right] + (\frac{\nu_1}{3}+\nu_2)\frac{\partial div\vec{V_1}}{r\partial \varphi};$$

$$\frac{\partial p_1}{\partial t} + w\frac{\partial p_1}{\partial r} + c^2\rho(\frac{1}{r}\frac{\partial rV_{1r}}{\partial r} + \frac{1}{r}\frac{\partial V_{1\varphi}}{\partial \varphi}) = 0;$$

$$div\vec{V_1} = \frac{1}{r}\frac{\partial(rV_{1r})}{\partial r} + \frac{1}{r}\frac{\partial V_{1\varphi}}{\partial \varphi} \qquad (2)$$

In (2), $\nu_1$ and $\nu_2$ are the first and second (volume) constant coefficients of kinematic viscosity; the perturbation fields are denoted by the subscript 1; $c^2 = (\partial p/\partial \rho)_s = \gamma p/\rho$ is the square of the speed of sound in the region behind the shock wave front where the Mach number $M = |w|/c < 1$.

In (2), the equation for pressure is obtained from the continuity equation and from the equation for entropy perturbation ($s = \ln \frac{p}{\rho^\gamma}; s \to s + s_1; p \to p + p_1; \rho \to \rho + \rho_1;$ $p_1 \ll p; \rho_1 \ll \rho; s_1 \ll s$):

$$\frac{\partial \rho_1}{\partial e} + w \frac{\partial \rho_1}{\partial r} + \rho \, div \vec{V}_1 = 0 \qquad (3)$$

$$\frac{\partial s_1}{\partial t} + w \frac{\partial s_1}{\partial r} = 0;$$
$$s_1 = \frac{1}{p}(p_1 - c^2 \rho_1) \qquad (4)$$

In (4), for simplicity, an example of the dependence of entropy perturbations on perturbations of the density and pressure of a polytropic gas is given. In general, for an arbitrary equation of state, the entropy perturbation can also be expressed in terms of density perturbation and pressure in the form of a known relation [10] (see the problem for paragraph 82 in [10])

$s_1 = \frac{1}{c^2}\left(\frac{\partial s}{\partial \rho}\right)_{p=const}(\rho_1 c^2 - p_1); c^2 = \left(\frac{\partial p}{\partial \rho}\right)_{s=const}$. In contrast to the case of a plane shock wave, it is not convenient to divide perturbations into acoustic-type perturbations and entropy-type perturbations, since Equation (4) no longer has a solution in the form of a plane traveling wave as in [10]-[12].

Equations (2) provide a closed description of the evolution of perturbations for pressure and velocity. Thus, a description can also be obtained for density and entropy perturbations based on the solution of System (2) using the continuity equation (3) and the definition of entropy perturbations from (4), or, in the general case, from the equation of state for an arbitrary medium.

For this reason it is also not necessary to make a special distinction between small hydrodynamics perturbations in (2)-(4) according to acoustic and entropy types as is traditionally made for plane shock waves in an ideal inviscid medium [10]-[12].

System (2) must be considered together with the boundary conditions on the perturbation (of the near-stationary cylindrical shock surface with $R_{s0} \approx const;$) surface which is defined by Function (1). It is possible to neglect the influence of viscosity on the boundary conditions in the limit of small viscosity (or large Reynolds numbers) [14]. Then, for the tangential and normal unit vectors to this surface, we have (see also [11, 12]):

$$\vec{t} = (t_r, t_\varphi) = (-g_\varphi / R_{s0}; -1) / \sqrt{1 + g_\varphi^2 / R_{s0}^2};$$
$$\vec{n} = (n_r, n_\varphi) = (1; -g_\varphi / R_{s0}) / \sqrt{1 + g_\varphi^2 / R_{s0}^2}; \qquad (5)$$
$$g_\varphi = \frac{\partial g(\varphi, t)}{\partial \varphi}$$

In the linear approximation, the nonlinear terms in (5) can be neglected. From the condition of continuity of the tangential component of the velocity field at the perturbed shock front, it follows that the scalar product of the vector $\vec{t}$ with the velocity vectors on both sides of the shock front has the form: $(w + V_{1r}, V_{1\varphi}) \circ \vec{t} = (w_0, 0) \circ \vec{t}; for: w_0 = U_0 - D = -D \approx const$.

Let's write this scalar product in more detail in the form $(w + V_{1r})t_r + V_{1\varphi}t_\varphi = w_0 t_r$. We further use (5) under the assumption of smallness of perturbations $g_\varphi \ll 1$, which allows us to discard nonlinear terms that are included under the sign of the square root when $t_r \approx -g_\varphi / R_{s0}; t_\varphi \approx -1$. After discarding the term $V_{1r} t_r$ that is also quadratic in the perturbation amplitude, we obtain the boundary condition in a linear approximation in terms of the perturbation amplitudes:

$$V_{1\varphi} = g_\varphi (w_0 - w) / R_{s0} \qquad (6)$$

Similarly, from the boundary condition for the normal component of the velocity field (which determines the difference between the scalar product of the normal vector $\vec{n}$ with the velocity vectors on both sides of the shock front), we obtain the relation

$(w + V_{1r}, V_{1\varphi}) \circ \vec{n} - (w_0, 0) \circ \vec{n} = w - w_0 + w_1$.

In this case, the value of the disturbance $w_1$ is determined from the well-known representation [10], [12] for $w - w_0 = \sqrt{(p - p_0)(1/\rho_0 - 1/\rho)}$, where $\delta = \rho/\rho_0 > 1$ using $p \to p + p_1; \rho \to \rho + \rho_1$, decomposing into a Taylor series in the limit $p_1 \ll p; \rho_1 \ll \rho$ and dropping the quadratic terms in the perturbation of density and pressure. Then the boundary condition for the normal component of the velocity field leads to the equation (see the same equation in [11, 12] and [14]):

$$V_{1r} = \frac{(w - w_0)}{2} \left( \frac{p_1}{p - p_0} + \frac{\rho_1}{\rho^2 (1/\rho_0 - 1/\rho)} \right) \qquad (7)$$

Let us determine the relationship of the perturbation of density and pressure in (7). In this connection, in [10]-[12], it is proposed to use the relation between density and pressure perturbations of the Hugoniot shock curve in the form of $p_1 = \left(\frac{dp}{d(1/\rho)}\right)_H \left(-\frac{\rho_1}{\rho^2}\right)$ or:

$$\rho_1 = -p_1 h \frac{\delta^2}{w_0^2};$$
$$h = j^2 \left( \frac{d(1/\rho)}{dp} \right)_H ; j^2 = \frac{p - p_0}{1/\rho_0 - 1/\rho} \qquad (8)$$

To find the equation that determines Function (1), we use the equality determining the perturbation of the velocity of the shock wave in the form $D_1 = \partial g / \partial t$ and the well-known expression [10, 12] $w_0^2 = D^2 = \frac{(p - p_0)\delta}{\rho_0(\delta - 1)}$. To get the expression $D_1$, it is necessary to make a replacement on the left side of the equality $D \to D + D_1$. Using $p \to p + p_1; \rho \to \rho + \rho_1$ and decomposing into the Taylor series in the limit $p_1 \ll p; \rho_1 \ll \rho$ after dropping the quadratic terms in the perturbation of density and pressure gives the equation (see also [12, 14]):

$$\frac{\partial g}{\partial t} = -\frac{w_0}{2}(\frac{p_1}{p-p_0} - \frac{\rho_1}{\rho^2(1/\rho_0 - 1/\rho)}) \qquad (9)$$

As in (7), in (9) the relation between density and pressure perturbations is given by Relation (8).

Let us look for a solution of Equations (2) with boundary conditions (6), (7) and Relations (8), (9) as

$$g(\varphi, t) = \sum_{n=0}^{N} \bar{g}_n \exp(i(k_n \varphi - \omega_n t));$$

$$(V_{1r}, V_{1\varphi}, p_1) = \sum_{n=0}^{N} (\bar{V}_{nr}, \bar{V}_{n\varphi}, \bar{p}_n) K_n(l_n r) \exp(i(k_n \varphi - \omega_n t)) \qquad (10)$$

In (10), $K_n(l_n r)$ is the MacDonald function (a modified Bessel's function of the second kind) of n-order that gives zero boundary conditions for perturbations at infinity $r \to \infty$. In (10), the values of the longitudinal wave numbers $l_n$ are always positive ($l_n > 0$); this is the result of the cylindrical symmetry. A negative value of $l_n$ is impossible in principle as is a negative value of the radial coordinate $r$. On the other hand, cylindrical symmetry is breaking for the plane SW where a new, chiral, symmetry arises (that is, an independence of the problem with respect to the direction of the SW velocity – whether from left to right or vice versa). Indeed, consider, as in [12,14], the case whereby the front of a plane SW moves from left to right in the positive direction of axis $x > 0$, the perturbations existing only after the front of the plane SW (in the system where the shock front is located at $x = 0$) in the region $x < 0$ and for zero boundary condition at $x \to -\infty$. In this case, it is necessary to consider only perturbations which are proportional to $\exp(ixl); l = il_1; l_1 < 0$ [12], [14]. Thus, for a plane shock wave, the condition that the longitudinal wave number is negative, i.e., $l_1 < 0$, is one of the (two) necessary conditions for the instability of the shock front.. Positive values are possible in principle, but they must be excluded by the zero boundary condition at infinite $x \to -\infty$. In this regard, it is also important to note the difference between the case of the converging SW and the case of the diverging cylindrical SW where the compression region after the front of SW is always finite; therefore the representation (8) for the case of a diverging SW must be changed on the condition at $r = 0$, or at the moving piston boundary.

Further, for simplicity, we will use the representation as given in Equations (10), in which we will limit ourselves to considering only one member of the series with a number $n = 1$. The choice of this number is related to the relative simplicity of the form of recurrent relations for modified Bessel functions in this case. However, we will not assign an index for the frequency and wave numbers used in the definitions given in Equation (10).

For an arbitrary value $n$, a similar consideration can also be obtained by analogy with the case considered below, when $n = 1$. For an arbitrary value $n$, it is indeed necessary to use more cumbersome considerations that use general recurrence relations.

For example, the ratio $\frac{1}{z}\frac{d}{dz}(zK_n(z)) = -K_{n-1}(z) - (n-1)K_n(z)/z$ must be applied when determining the divergence of the velocity field. When taking into account the viscosity and the corresponding term in the Laplace operator, in general, a recurrence relation of the form

$\frac{1}{z}\frac{d}{dz}(z\frac{dK_n}{dz}) - \frac{K_n}{z^2} = K_{n-2} + 2(n-1)\frac{K_{n-1}}{z} + (n^2-1)\frac{K_n}{z^2}$ should be used where for the case of integers $n$ the relation holds: $K_n(z) = K_{-n}(z)$.

Note that, with accuracy up to first order, when substituting Equations (10) into the boundary conditions (6), (7), relations (8), (9) also should be considered on the unperturbed surface of the shock front.

To obtain the solution of the system (6)-(9), the unknown function $g$ can be excluded from (6) and (9). To do this, the left and right parts of Equation (6) must be differentiated by time $t$, and the left and right parts of Equation (9) must be differentiated by the angular variable $\varphi$. Then substitute the representation $\dfrac{\partial^2 g}{\partial t \partial \varphi} = \dfrac{R_{s0}}{w_0 - w} \dfrac{\partial V_{1\varphi}}{\partial t}$ obtained from (6) after time differentiation into the left part of (9) obtained after differentiation with respect to the angular variable. As a result, from (6) and (9), taking into account (8), we obtain the equation:

$$\frac{\partial V_{1\varphi}}{\partial t} = -\frac{w_0(w_0 - w)(1+h)}{2(p - p_0)R_{s0}} \frac{\partial p_1}{\partial \varphi} \quad (11)$$

From (7), also taking into account (8), we obtain the equation:

$$V_{1r} = \frac{(w - w_0)(1-h)}{2(p - p_0)} p_1 \quad (12)$$

The system of Equations (11) and (12) contains three unknown functions, that is, perturbations of the two components of the velocity field and pressure perturbations, which must be considered at the boundary of the studied region coinciding with the position of the undisturbed shock front at the value of the radial coordinate $r = R_{s0}$.

To close the system (11), (12), we use the equation for pressure perturbations from (2), considering it exactly on the specified boundary surface in the form:

$$\left(\frac{\partial p_1}{\partial t}\right)_{r=R_{s0}} + w\left(\frac{\partial p_1}{\partial r}\right)_{r=R_{s0}} + c^2 \rho \left(\frac{1}{r}\frac{\partial r V_{1r}}{\partial r} + \frac{1}{r}\frac{\partial V_{1\varphi}}{\partial \varphi}\right)_{r=R_{s0}} = 0 \quad (13)$$

After substitution of (10) (only for the case with $n=1$) in the resulting system of equations, from (5) and (6), a homogeneous system of equations follows:

$$\bar{V}_{1r} = \frac{(1-h)}{2\rho_0 |w_0|} \bar{p}_1;$$

$$\bar{V}_{1\varphi} = \frac{(1+h)k}{2R_{s0}\omega\rho_0} \bar{p}_1$$

$$\bar{p}_1 = \frac{c^2\rho}{(\omega - wl_1)} (\frac{k}{R_{s0}} \bar{V}_{1\varphi} + \bar{V}_{1r} l_2)$$

$$l_1 \equiv il(\frac{K_0(lR_{s0})}{K_1(lR_{s0})} + \frac{1}{lR_{s0}}); l_2 \equiv il\frac{K_0(lR_{s0})}{K_1(lR_{s0})}$$

(14)

In (14), $K_1$ and $K_0$ (modified Bessel functions of the first and zero'th order, respectively, abbreviated-- $l_1$ and $l_2$) as well as the notation for expressions containing the radial wave number $l$ (which is positive and real) are introduced.

From the condition of solvability of the system (14) we obtain the following dispersion equation which gives the generalization of Equation (2.15) in [14] on the cylindrical case in the limit of zero viscosity:

$$\omega + Mcl_1(1 - \frac{(1-h)l_2}{2M^2 l_1}) = \frac{(1+h)k^2 c^2 \delta}{2R_{s0}^2 \omega}$$

(15)

In the limit $lR_{s0} \to \infty$ Equation (15) has exactly the same form as in the case of a plane SW in the limit of high Reynolds numbers. See (2.15) in [14] where $a_2 \to 1-h; a_1 \to 1+h$ in the limit $\text{Re}_\lambda = |w_0|\lambda/\nu \gg 1$; $\lambda$ is the width of the plane SW which is connected with viscosity $\nu$. This similarity occurs because in this limit $K_0(lR_{s0})/K_1(lR_{s0}) \to 1$. It is only necessary to replace $k/R_{s0}$ in (15) with the wave number $k$ of 2D corrugation perturbations, which is not dimensionless as $k$ is in (15). But even in this limit, there is a substantial difference in the determination of the longitudinal wave number $l$ between the converging cylindrical SW in (15) and the plane SW, as it is described above. Indeed, in (15) the value $l$ is always positive.

Dispersion Equation (15) must be considered together with the dispersion equation which is defined after substitution of (10) into (2) and integration of all equations $\int_0^\infty drr^3$ (when relations $\int_0^\infty dr r^{\mu-1} K_\nu(lr) = \frac{2^{\mu-2}}{l^\mu} \Gamma(\frac{\mu+\nu}{2})\Gamma(\frac{\mu-\nu}{2})$ are taken into account, where $\Gamma$ is the gamma function):

$$A_1 A_2 = k^2 l^2 A_3. \qquad (16)$$

Where
$$A_1 = \Omega + 2Mcl + \frac{3\nu_1 \pi d^2}{2}; \Omega = i\omega \frac{3\pi}{2} - 6Mcl;$$
$$A_2 = \Omega(\Omega + \frac{3\pi d^2}{2}(\frac{4}{3}\nu_1 + \nu_2)) - 24c^2 l^2;$$
$$A_3 = \pi(\frac{4}{3}\nu_1 + \nu_2)\Omega^2 - 4\Omega c^2(1-\varepsilon) - 2\pi c^2 l^2 \left(\frac{17}{3}\nu_1 + 2\nu_2\right);$$
$$\varepsilon = \frac{\pi^2 l^2}{16c^2}\left(\frac{13}{3}\nu_1 + \nu_2\right).$$

The dispersion equation given by (16) is derived from (2) and (10) as the weak solution of (2), while the dispersion equation given by (15) is the result of the strong solution which is obtained exactly from boundary conditions (6), (7) and (9) for representations (8) and (10).

## 2. Instability of a shock wave for 1-D perturbations.

The instability of a shock is possible when, in (15) and (16), for the positive real value of the radial wave number $l > 0$, a solution with $\omega_1 = \text{Im}\,\omega > 0; \omega = i\omega_1$ can be obtained.

Let us consider the solutions of dispersion equations (15) and (16) in the limit $k \to 0$ which corresponds to the case of one-dimensional (1D) perturbations. From (16) in this case it is necessary that $A_1 A_2 = 0$. For the case when $A_1 = 0$, from (16) it is possible to obtain the representation:

$$\omega_1 = v_1 l^2 - \frac{8Mcl}{3\pi} \qquad (17)$$

Equation (15) in the limit $k \to 0$ may be represented in the form:

$$\omega_1 = Mcl \left( \frac{K_0(lR_{s0})}{K_1(lR_{s0})} + \frac{1}{lR_{s0}} \right) \left[ \frac{(1-h)}{2M^2 \left( 1 + \frac{K_1(lR_{s0})}{lR_{s0} K_0(lR_{s0})} \right)} - 1 \right] \equiv Q(lR_{s0}) \qquad (18)$$

From (17) and (18) it is possible to obtain an equation for the wave number:

$$l(v_1 l - \frac{8Mc}{3\pi}) = Q(lR_{s0}) \qquad (19)$$

Let us consider Equation (19) in the limit $lR_{s0} \gg 1$ when one can write

$\frac{K_1(lR_{s0})}{K_0(lR_{s0})} \approx 1 + \frac{1}{2lR_{s0}} + O(1/l^2 R_{s0}^2)$. In this limit, from (19) the equation for $l$ is rewritten:

$$\begin{aligned} & l^2 - l_S l + \frac{Mc}{2v_1 R_{s0}} (1 + \frac{1-h}{2M^2}) = 0. \\ \text{Where} \\ & l_S = \frac{Mc}{v_1} \left( \frac{1-h}{2M^2} - 1 + \frac{8}{3\pi} \right) \end{aligned} \qquad (20)$$

For the solution of (20) it is possible to obtain the representation:

$$\begin{aligned} & l = l_\pm = \frac{l_S}{2} \left[ 1 \pm \sqrt{1 - \frac{4M(1 + 2M^2 - h)}{\text{Re}(1 - h - 2M^2(1 - 8/3\pi))^2}} \right]. \\ \text{where} \\ & \text{Re} = \frac{R_{s0} c}{v_1} \end{aligned} \qquad (21)$$

For the limit of large Reynolds numbers $\text{Re} \gg 1/M$ in (21), one of the two solutions (21) is equal to the value:

$$l = l_+ \approx l_S (1 + O(1/M \text{Re})) \tag{22}$$

From (22) and (17) it is possible to obtain an expression for the exponential growth rate of the perturbations:

$$\omega_1 = \frac{c^2}{2\nu_1}(1 - 2M^2 - h)\left(\frac{1-h}{2M^2} - 1 + \frac{8}{3\pi}\right)(1 + O(1/M \text{Re})) \tag{23}$$

The necessary and sufficient condition for the instability of the converging cylindrical SW with $\omega_1 = \text{Im}\,\omega > 0$ in (23) (when also $l = l_S > 0$ in (22), see (20)) is:

$$h < h_1 \equiv 1 - 2M^2 \tag{24}$$

It is important that even in the limit $lR_{s0} \gg 1$ the condition (24) is substantially different from the conditions of instability for a plane SW in an ideal medium in the D'yakov theory [10]-[12], where it is necessary that $h > 1 + 2M$ or $h < -1$ for 2D perturbations. Moreover, the condition of instability of 1D and 2D perturbations for a plane SW in a viscous medium is the inequality $1 - 2M^2 < h < 1$ [14] which is opposite to the condition (24).

An additional solution of (21) and (17) also exists in the form (when $\text{Re} \gg 1/M$ in (21)):

$$l = l_- = \frac{(1 + 2M^2 - h)}{R_{s0}\left(1 - h - 2M^2(1 - \frac{8}{3\pi})\right)} \tag{25}$$

$$\omega_1 = \nu_1 l_-^2 \left(1 - \frac{8M \text{Re}\left(1 - h - 2M^2(1 - \frac{8}{3\pi})\right)}{3\pi(1 + 2M^2 - h)}\right) \tag{26}$$

From (26) it is possible to obtain the instability condition in the form:

$$h_3 \equiv h_2 - \frac{4M^2(1-4/3\pi)}{\frac{8M\operatorname{Re}}{3\pi}-1} < h < h_2 \equiv 1 - 2M^2(1-\frac{8}{3\pi});  \quad (27)$$

$$M\operatorname{Re} > 3\pi/8$$

In the limit $\operatorname{Re} >> 1/M$, in (26) the value $\omega_1 \cong \nu_1/R_{s0}^2$ is much smaller than in (23). So, the dominant condition of instability is condition (24), but the estimation $\omega_1 \cong \nu_1/R_{s0}^2$ gives the tendency for the increasing of the exponential instability growth rate $\omega_1$ when the radius of the converging shock is decreasing. Moreover, condition (27) restricts the possible values of $h$ from above to a lesser extent, since $h_1 < h_2$. Therefore, the case is possible whereby the instability condition (24) is impossible, but the condition (27) is fulfilled for $h_1 < h_3 < h < h_2$.

It must be also noted that in the limit $lR_{s0} << 1$ when $K_0(lR_{s0})/K_1(lR_{s0}) \approx lR_{s0}\ln(2/lR_{s0})$ in (18), instability is impossible. This is due to the fact that, in this case, from (17) and (19) follows $l = \frac{Mc}{3\pi\nu_1}\left(1 \pm \sqrt{1-9\pi^2/M\operatorname{Re}}\right), \omega_1 = -Mc/R_{s0} < 0$.

For the case with $A_2 = 0$, shock instability is possible only in the limit $lR_{s0} >> 1$ when the condition of instability is the same as (24), because for the case $\operatorname{Re}/lR_{s0} = \frac{c}{\nu_1 l} << 1$ it is possible to obtain the expressions:

$$l = \frac{Mc}{(\frac{4}{3}\nu_1+\nu_2)}\left(\frac{1-h}{2M^2}+\frac{4}{\pi}-1\right);$$
$$\omega_1 = \frac{c^2(1-2M^2-h)}{2(\frac{4}{3}\nu_1+\nu_2)}\left(\frac{1-h}{2M^2}+\frac{4}{\pi}-1\right) \quad (28)$$

From (28) it is possible to again obtain the condition of instability in the form of (24).

On the other hand, in the limit $lR_{s0} << 1; k >> 1$, instability is impossible, as for the case $A_1 = 0$.

### 3. Instability of a shock wave for 2-D perturbations

Let us consider a 2D perturbation with $k \neq 0$ in the limit $k >> 1$ in (15) and (16). Then, from (16) in this limit when $A_3 = 0$ it is possible to obtain a representation of (16) in the form:

$$\Omega^2 - 8\Omega \frac{c^2(1-\varepsilon)}{\pi(v_2 + v_1/3)} - 4c^2 l^2 \left(2 + \frac{5v_1}{v_2 + v_1/3}\right) = 0;$$

$$\varepsilon = \frac{\pi^2 l^2}{16c^2}\left(\frac{4}{3}v_1 + v_2\right)\left(\frac{13}{3}v_1 + v_2\right)$$

(29)

The solution of (29) in the limit $\varepsilon << 1$ has the representation:

$$\Omega = \Omega_\pm = \frac{4c^2}{\pi(v_2 + v_1/3)}\left(1 \pm \sqrt{1 + \frac{l^2\pi^2(v_2 + v_1/3)(2v_2 + 17v_1/3)}{4c^2}}\right) \quad (30)$$

In (30), the case $\Omega = \Omega_+$ corresponds to the stability of the shock wave front, because $\omega_1 < 0$. When $\Omega = \Omega_-$ in the limit $lv_1/c \cong \varepsilon << 1$ it is possible to obtain the expression

$\omega_1 \equiv \omega_{1\Omega} \approx -\frac{4lMc}{\pi} + \frac{l^2}{3}(2v_2 + 17v_1/3)$ which gives the possibility of shock instability.

From the dispersion equation (15), for a finite value of the dimensionless wave number $k > 0$ in the limit $R_{s0} l >> 1$, we can obtain a solution in the form:

$$\omega_1 = \omega_{1\pm} = \frac{cl(1 - 2M^2 - h)}{4M}\left(1 \pm \sqrt{1 - \frac{8M^2 k^2 \delta}{l^2 R_{s0}^2 (1 - 2M^2 - h)^2}}\right) \quad (31)$$

From $\Omega = \Omega_-$ in (30), in the limit $\varepsilon \cong l^2 v_1^2/c^2 << 1$ and in the case $R_{s0} l >> k >> 1$, when in (31) $\omega = \omega_+ \approx \frac{cl(1 - 2M^2 - h)}{2M}(1 - O(\frac{k^2}{l^2 R_{s0}^2}))$, it is possible to obtain the expressions:

$$\omega_1 = \frac{3c^2\left(1-2M^2-h\right)\left(1-O\left(\frac{k^2}{l^2 R_{s0}^2}\right)\right)}{4M^2(2\nu_2+17\nu_1/3)}\left[1-h+2M^2\left(\frac{4}{\pi}-1\right)\right]; \quad (32)$$

$$l = \frac{3c}{2M(2\nu_2+17\nu_1/3)}\left[1-h+2M^2\left(\frac{4}{\pi}-1\right)\right]$$

According to (32), the instability condition in the two-dimensional case coincides with the instability condition for one-dimensional perturbations given in (24). In this case, the difference between the exponential increments of one-dimensional perturbations in (23) and two-dimensional perturbations in (31) is insignificant in cases where the volume (second) viscosity coefficient is small compared to the shear viscosity coefficient. However, in the opposite limit $\nu_2 \gg \nu_1$, which is typical for systems far from equilibrium (see [10]), the rate of increase of one-dimensional perturbations can significantly exceed the rate of increase of two-dimensional perturbations.

In the case when $\omega_1 = \omega_{1-}$ in (31) in the limit $\varepsilon \ll 1$, we get $\omega_1 \approx \frac{cMk^2\delta}{lR_{s0}^2(1-2M^2-h)}$. In this case, taking into account the corresponding representation following from (30) in the case of $\Omega = \Omega_-$ we obtain for the longitudinal wave number of perturbations of the shock wave front and for the increment of perturbation growth over time:

$$l = \frac{12Mc}{\pi(2\nu_2+17\nu_1/3)}(1+O(k^2/l^2 R_{s0}^2));$$

$$\omega_1 = \frac{\pi k^2 \delta(2\nu_2+17\nu_1/3)}{12 R_{s0}^2(1-2M^2-h)} \quad (33)$$

It follows from (33) that the condition $\omega_1 > 0$ of instability of the shock wave front with respect to two-dimensional perturbations coincides with the condition (24) obtained for the case of one-dimensional perturbations. In this case, according to (33) and (24), the growth rate of two-dimensional perturbations will exceed the growth rate of one-dimensional perturbations if the condition is met $R_{s0} < R_{th1} = \dfrac{k}{2c\sqrt{3}}\sqrt{\nu_1(2\nu_2 + 17\nu_1/3)}$.

A similar restriction on the value of the unperturbed radius of the shock wave is obtained from comparing exponential growth of two-dimensional perturbations corresponding to the two different solutions (31) presented in (32) and (33). Indeed, the exponential growth rate of perturbations in (33) becomes greater than the exponent in (32) only for sufficiently small values of the unperturbed radius of a converging cylindrical shock wave:

$$R_{s0} < R_{th2} = \dfrac{kM}{3c(1-2M^2-h)}\left[\dfrac{\pi\delta(2\nu_2 + 17\nu_1/3)(\nu_2 + 17\nu_1/3)}{1-h+2M^2(\dfrac{4}{\pi}-1)}\right]^{1/2} \quad (34)$$

Thus, when (34) takes place, perturbations in the case (33) become more dominant than for (32) when a sufficiently small shock wave radius is attained. The threshold value (34) is proportional to the dimensionless value of the azimuthal number. Therefore, with an increase in this wave number the threshold value of the shock wave radius increases, at which the instability of relatively small two-dimensional perturbations is most pronounced in regime (33).

Note that in the limit $lR_{s0} \ll 1$ of large wavelengths of longitudinal perturbations for two-dimensional perturbations, in contrast to the case of one-dimensional perturbations, the instability of the shock wave front is possible under the condition

$$h < -1 \quad (35)$$

Condition (35) coincides with the condition of instability of plane shock waves obtained in the D'yakov theory without taking into account the effects of viscosity. We show that condition (35) is only necessary, but not sufficient for the realization of instability of cylindrical shock waves. Indeed, in the limit $lR_{s0} \ll 1$ of (16) (taking into account the representation $K_0(z)/K_1(z) \approx z\ln(2/z); z = lR_{s0} \ll 1$), we obtain for the increment of the exponential evolution of perturbations:

$$\omega_1 \equiv \omega_{1K} = \frac{c}{R_{s0}}\left(-\frac{M}{2} + \sqrt{\frac{M^2}{4} - \frac{(1+h)k^2\delta}{2}}\right) \qquad (36)$$

It follows from (36) that the necessary instability condition $\omega_1 > 0$ has the form (35). However, in order to realize the instability, it is required that in addition to (36), the condition $\omega_1 = \omega_{1\Omega} > 0$ is fulfilled, where $\omega_{1\Omega}$, as noted above, corresponds to the solution $\Omega_-$ of Equation (30) in the limit $\varepsilon \ll 1$. In this case, the equality $\omega_{1K} = \omega_{1\Omega}$ is also required, from which, under condition (35), the dimensionless transverse wave number of perturbations is determined in the form:

$$k = k_\Omega(l) = \left[\frac{2\omega_{1\Omega} R_{s0}\left(\frac{\omega_{1\Omega} R_{s0}}{c} + M\right)}{c\delta(|h|-1)}\right]^{1/2} \quad ; h = -|h|$$

$$\omega_{1\Omega} = \frac{l^2}{3}(2\nu_2 + 17\nu_1/3)\left(1 - \frac{l_{th}}{l}\right); \qquad (37)$$

$$l > l_{th} = \frac{12Mc}{\pi(2\nu_2 + 17\nu_1/3)}$$

Conditions (35) and (37) are necessary and sufficient for realizing the instability of the cylindrical shock wave front with respect to only two-dimensional perturbations. As noted above, in the limit $lR_{s0} \ll 1$, one-dimensional perturbations at $k = 0$ cannot lead to instability.

Thus, from (33) for the 2D perturbations and from (23) and (28) for the 1D perturbations, it is easy to see that the value of the growth rate for 1D perturbations is larger than for the case of 2D perturbations of a cylindrical converging SW in the limit of large Reynolds numbers, or for small viscosity. On the other hand, the condition of instability (24) is the same for 1D and 2D cases. In the limit of long-wave longitudinal perturbations, as follows from (37), the realization of instability is possible only for two-dimensional perturbations. In this case, the front of a converging cylindrical shock wave is stable relative to one-dimensional longitudinal perturbations.

## 4. Discussion and comparison with experiment

The instability of the shock wave front was studied experimentally and numerically in [15-23]. For example, in [16] the stability of strong shock waves at speeds from 3 to 18 km/s in argon and carbon dioxide was studied in a free-piston shock tube using time-resolved interferometry. In [16] it was suggested that the linearized perturbation analysis of D'yakov theory [11] is not accurate for the case of real shock tube flow. The shock wave front in the experiment [16] has a non-zero curvature due to the influence of the solid side boundaries of the shock tube. In this case, the shock wave still differs from the plane shock wave considered in [11]. In this regard, it makes sense to compare the instability conditions (24) obtained above with the data obtained in the experiment [16]. We show that the instability condition (24), although obtained in the framework of linear theory, still agrees quite well with the data of shock wave instability observations in [16].

In [16], the instability of the shock wave in argon was established experimentally in the range of shock wave velocities from 10 to 11 km/s. For a shock wave in carbon dioxide, instability is found in the range of shock wave velocities between 5 and 6 km/s, as well as near 3.5 km/s. Figure 8 in [16] shows the dependences on the shock wave velocity for the D'yakov parameter h and for the threshold values of this parameter $h_{D\max} = \dfrac{1 - M^2 - M^2\delta}{1 - M^2 + M^2\delta}$, which defines the upper bound of stability in the theory [11]. In [11], it is found that the plane shock wave front is stable in the linear theory, if the condition $-1 \leq h \leq h_{D\max}$ is met. In this case, it turns out that the curve of the dependence $h$ on the velocity of the shock wave, at the values of the velocity corresponding to the instability observed experimentally in [16], is always below the curve $h_{D\max}$ (see Fig.8 in [16]). Therefore, the instability of the shock wave detected in [16] is realized precisely under the condition $h < h_{D\max}$; that condition in the theory of [11], on the contrary, corresponds to the stability of the shock wave.

Due to the inequality $\delta = w_0 / w > 1$, the inequality $h_{D\max} < h_1 = 1 - 2M^2$ is always valid, where $h_1$ defines the upper limit in the condition of instability (24). Thus, the range of variation of the D'yakov parameter $h < h_{D\max} < h_1$, in which the instability of the shock waves was experimentally observed in [16], corresponds to the condition of instability (24).

The realization of the instability of the shock wave was observed in [16] when the value $h$ was positive and varied from zero to 0.05, in the range of speed change shock waves from 10 to 11 km/s in argon, and when the speed of the shock wave was from 5 to 6 km/s in carbon dioxide.

At the same time, the shock wave instability observed in [16] near the shock wave velocity $D \approx 3.5$ km/s was realized even at a negative parameter value $h \approx -0.025$. All these values of the D'yakov parameter $h$ satisfy the instability condition (24).

Thus, instability was observed in [16] in a narrow range of changes in parameter $-0.025 \leq h \leq 0.05$; therefore it follows from condition (24) that instability can only occur if in addition the inequality $M \leq 0.689$ holds. For example, for the case when the velocity of the shock wave in argon is equal to $D \approx 10.5$ km/s, and the compression $\delta \approx 13.33$, which corresponds to the observed instability of the shock wave at the parameter value $h_{D\max} \approx 0.05$, is equal to $M \approx 0.245$ in [16]. This value is consistent with the inequality noted above, which follows from the instability condition (24).

In [16], it is noted that the instability of the shock wave front is most pronounced in the region of positive values of the D'yakov parameter $h > 0$, which is realized due to dissipative processes associated with completion of first ionization or dissociation. The mechanism of instability of a converging cylindrical shock wave considered in this paper is also dissipative, taking into account the effect of viscosity.

The important role of instability formation with respect to one-dimensional purely radial perturbations is shown, which can then contribute to the formation of corresponding azimuthally inhomogeneous perturbations. They first appear behind the shock wave front by analogy with instability in the boundary layer [14] and then form increasing perturbations of the shock wave front itself. This mechanism of instability allows us to interpret the known data of experimental observations of shock wave instability associated with various forms of dissipative processes from a single position.

Consider, for example, the results of the experimental observations in [17] and [18]. In [18], for the converging shock waves into different gases such as argon and xenon, an axial laser probing images shows the shock fronts to be smooth and azimuthally symmetric, with instabilities developing only downstream from each shock. Instabilities behind the first shock develop as the shock propagates (see Fig. 10a and b in [18]). For example, the Ar image (Fig. 10a) shows 50 μm-wide fingers extending 250 μm outward radially behind the first shock front by the time it reaches a radius of 1.5 mm.

The observations of well-defined instabilities are in contrast with minimal perturbations seen behind the first shock at earlier times. As stated in [18], the cause of the instabilities in the relaxation region is not yet clear, though qualitatively they appear more pronounced in Xe than in Ar. And hence this could be related to the strength of radiative losses and/or the amount of compression at the shock front [18]. Moreover, the shock structure changed dramatically in the same experiments with nitrogen ($N_2$) and as stated in [18], the reason for such change in the shape of the shock is not known at present. In [18], in the experiments performed so far with nitrogen, the azimuthal symmetry was poor and was characterized by a $k = 5 \div 6$ (near hexagonal) structure. Despite this, the shock front itself remains very thin and well defined. In [18], for nitrogen a large degree of dissociation is expected to occur across the shock front, or even in the pre-ionised region ahead of the shock, which, in turn, could provide a seed for instabilities to grow.

We can therefore assume that the experimentally detected instability of the shock wave front in [16] and [18] had the same dissipative mechanism. Therefore, the conditions of dissipative instability of a converging cylindrical shock wave obtained in this work can be applied not only to analyze the results [16], as presented above, but also in connection with the data obtained in the experiment [18]. Since [18], in contrast to [16], does not provide the necessary data for such an analysis of the Hugoniot shock curve, it is not yet possible to conduct a quantitative comparison of the theory developed in this paper and the data from [18].

In this connection, we also note the results of experimental observation of the instability of the front of a converging cylindrical shock wave in air, obtained in [17] and allowing us to better understand the results of [18]. Indeed, [17] indicates a close relationship between the observed instability of the flow behind the shock wave front and the instability of the shock wave front itself, which is in accordance with the results of this work. At the same time, in [18] such a connection was established only in the case of instability of a converging cylindrical shock wave in nitrogen, and for argon and xenon in [18], the instability of the shock wave front itself was not detected, despite the presence of instability of the medium flow behind the shock wave front.

In connection with the analysis of the results of [15] and subsequent experimental studies of the stability of converging cylindrical shock waves, [17] notes the importance of the methods used to register non-homogeneities of the flow behind the shock wave front and the deformation of the shock wave front itself. "Since in previous experimental work the shadow or schlieren methods were mainly used, the non-uniformity of flow behind the converging cylindrical shock wave was not easily detectable. The present holographic interferometer was particularly useful in detecting the earliest onset of instability in the flow preceding deformation of the shock wave." [17].

In [17], it is noted (see Fig.11 in [17]) that when the shock wave front is unstable in air, its azimuthal symmetry, in particular, is characterized by an azimuthal wave number $k = 4$, which does not differ much from the above value $k = 5 \div 6$, obtained for an unstable converging cylindrical shock wave in nitrogen, according to [18]. In [17], the instability regimes characterized by the wave numbers $k = 2$ and $k = 8$ were also considered. In [17], a tendency to increase the deformation of the shock wave front with a decrease in the azimuthal wave number $k$ was established.

This is in agreement with the estimate of the exponential growth of perturbations in (32) obtained in this paper. The trend indicated in [17] (see Fig.13 in [17]), however, is realized only when the unperturbed values $R_{s0}$ of the shock wave radius are not too small. Thus, according to [17], at $R_{s0} \leq R_{thE} \approx 0.65$ mm an inverse trend occurs when the shock wave front deformations increase with growth. This is consistent with another instability mode (33), which, for a sufficiently small shock wave radius defined in (34), actually begins to dominate in comparison with the instability mode (32).

To quantify the threshold radius $R_{th2}$ determined from (34), we take into account the well-known fact [10] that for strongly non-equilibrium systems relaxing to equilibrium, the value of the second (or bulk) viscosity can much exceed the value of the shear viscosity when the inequality $\nu_2 \gg \nu_1$ is fulfilled in (34).

In this case, from (34) we obtain an estimate for the threshold radius $R_{th2} \cong \dfrac{k\nu_2}{3c}\sqrt{2\pi\delta}$ of the shock wave. We use the representation $\nu_2 = \varsigma/\rho = \tau(c^2 - c_0^2)$ for the second (volume) viscosity coefficient given in [10] (see Formula (81.7) in [10]), corresponding to the low-frequency limit or small values of relaxation times $\tau$.

As a result, we get an estimate $R_{th2} \cong \tau c_0 k \dfrac{\sqrt{2\pi\delta}(z^2-1)}{3z}; z = c/c_0 = \sqrt{p/p_0\delta}$. For the values $p/p_0 \approx 200; \delta \approx 17$, we obtain that the values of the threshold radii $R_{th2} \approx R_{thE} = 0.65\,\text{mm}$, obtained from (34) and in [17], coincide if we take $k = 4;\ c_0 = 330\,\text{m/s}$ and $\tau = \lambda_s^2/\nu_1 \approx 0.46 \times 10^{-7}\,\text{s}$, where $\lambda_s \approx 8.3 \times 10^{-5}\,\text{cm}$ is the width of a shock wave front and $\nu_1 \approx 0.15\,\text{cm}^2/\text{s}$ is the kinematic viscosity coefficient for air.

Similar to the conclusions obtained in [17], the stability of a converging cylindrical shock wave in water was also studied in [19], where it was found that a noticeable instability of the shock front is observed only from those perturbations that occurred at the earliest stage of compression and had time to grow during compression.

As in [17], in [19] the growth rate of non-uniformities increases with their wave-length and, when the wavelength of non-uniformities becomes comparable to the distance from the converging axis (as for example in the case of dipole-shaped front of SW), they exhibit substantial growth, which considerably attenuates the energy accumulation in the vicinity of the converging axis. This means that the data of [19] on the stability of the shock wave in water also agree with the instability regime described in (32), where the exponential growth of perturbations also increases with a decrease in the azimuthal wave number $k$.

Note that, for example, for water, the characteristic time of change in the radius of the shock wave is of the order of $\tau_R \cong 10^{-8}\,\text{s}$, which is much longer than the characteristic time of exponential growth of perturbations $\tau_{\exp} \cong \left( \dfrac{c^2}{\nu_1} \right)^{-1} \approx 0.4 \times 10^{-12}$ s according to estimates (23), (28) and (32) (if we assume for water $c \cong c_0 \approx 1500\,\text{m/s}; \nu_1 \approx 10^{-2}\,\text{cm}^2/\text{s}$). This means that the quasi-stationary approximation $\varepsilon = \tau_{\exp}/\tau_R \ll 1$ considered above, which does not take into account the change in time of the undisturbed radius of the shock wave $R_{s0}(\varepsilon t) \approx const$, has a fairly wide range of applicability.

The given correspondence of the theory conclusions with the observational data [16]-[19] indicates the possibility of implementing the dissipative mechanism of shock wave front instability, which is also noted in [23].

We also note that the consideration carried out in this paper concerns only the perturbation mode (10) with the number n=1. A generalization to the case of accounting for modes with arbitrary numbers for analyzing the stability of converging cylindrical shock waves is of interest. Indeed, when considering the stability of converging spherical shock waves in [5], it was found that modes with sufficiently large numbers are stable. Therefore, setting a similar threshold mode number for converging cylindrical shock waves may be of practical significance.

## 5. Conclusions

The condition of exponential instability for the converging cylindrical SW is obtained only when viscosity is taken into account. This condition is more widely realizable than the condition for a plane SW dissipative instability which is obtained in [14]. Here only the case of small perturbations is considered in the quasi-stationary limit, when the characteristic time of the perturbation growth is much smaller than the characteristic the SW-radius change in time. In this case, the possibility of dissipative instability with respect to the one-dimensional perturbations that do not violate the cylindrical symmetry of the shock wave is shown, in contrast to previous studies that do not take into account the viscosity (see [6]). Moreover, due to the dissipative instability of the front of a converging cylindrical shock wave, a vortex flow may occur that has not only a radial, but also an azimuthal component of the velocity field of the material behind the shock wave front in the case of two-dimensional perturbations. As a result, it is impossible to obtain an ultra-high pressure mode when the shock converges to the symmetry axis under the conditions of dissipative instability obtained in this paper.        .

## Acknowledgments

I would like to thank Ya. E. Krasik for his attention to the work and discussions.

The work was supported by Israel Science Foundation, Grant number: 492/18

Data Availability Statements: **"The data that support the findings of this study are available from the corresponding author upon reasonable request."**